# Generating a Generic Fluent API in Java

Tomoki Nakamaru[a] and Shigeru Chiba[a]

a The University of Tokyo

**Abstract**

**Context:** Algorithms for generating a safe fluent API have been actively studied in recent years. A safe fluent API is a fluent API that reports incorrect chaining of the API methods as a type error to the API users. Although such a safety property improves the productivity of users, the construction of a safe fluent API is excessively complicated for developers. Therefore, generation algorithms are studied to reduce the development costs of a safe fluent API. Such studies may benefit many programmers, as a fluent API is a popular design in the real world.

**Inquiry:** The generation of a generic fluent API remains unaddressed. A generic fluent API refers to a fluent API that provides generic methods (those that contain type parameters in their definitions). The Stream API in Java is an example of a generic fluent API. Previous studies on safe fluent API generation have focused on the grammar classes that the algorithms can deal with for syntax checking. The key concept of such studies is using nested generics to represent a stack structure for the parser built on top of the type system. Within this context, the role of a type parameter has been limited to the internal representation of a stack element of that parser on the type system. Library developers cannot use type parameters to include a generic method in their API so that the semantic constraints for their API will be statically checked; for example, the type constraint on the items passed through a stream.

**Approach:** We propose an algorithm for generating a generic fluent API. Our translation algorithm is modeled as the construction of deterministic finite automaton (DFA) with type parameter information. Each state of the DFA holds information regarding which type parameters are already bound in that state. This information is used to identify whether a method invocation in a chain newly binds a type to a type parameter or refers to a previously bound type. This identification is necessary because a type parameter in a chain is bound at a particular method invocation, and the bound type is referenced in the following method invocations. Our algorithm constructs the DFA by analyzing the binding times of the type parameters and their propagation among the states in a DFA that is naively constructed using the given grammar.

**Knowledge:** With our algorithm, the generation of a safe fluent API can be introduced into practical use.

**Grounding:** We implemented a generator named Protocool to demonstrate our algorithm. Moreover, we generated several libraries using Protocool to demonstrate the ability and limitations of our algorithm.

**Importance:** Our algorithm can aid library developers to develop a generic fluent API, which is essential for bringing safe fluent API generation to the real world, as the use of type parameters is a common technique in library API design.

## 1 Introduction

Algorithms for generating a *safe* fluent API have been actively studied in recent years. A fluent API is an API that is designed to be used in the method-chaining style [12]. A safe fluent API is a fluent API that reports incorrect chaining of the API methods as a type error. The correct order of method invocations can be regarded as the *syntax* of a method chain. According to this viewpoint, the construction of a safe fluent API can be modeled as the construction of a parser on top of the type system. Several researchers have proposed algorithms that translate BNF-style method-chaining grammar into class definitions for a safe API [3, 16, 17, 23, 29]. As such construction is too complicated to carry out by hand, these algorithms reduce the development costs of safe APIs.

Recent research on safe fluent API generation has focused on the syntax-checking capability, which has been improved dramatically thanks to certain studies.

The study by Xu presented an algorithm to translate LL grammar into a safe fluent API [29]. Gil and Levy proposed an algorithm to translate LR grammar into a safe fluent API [16]. Furthermore, Gil and Roth proposed another algorithm for LR grammar to overcome the compilation time problem of the algorithm proposed in the literature [17]. In the study by Grigore [18], the Java type system was demonstrated as Turing complete, which indicates that the Java type system can even check context-sensitive grammar. Their key concept for checking context-free structures was using nested generics (parameterized types) to represent a stack structure on top of the type system. The role of a type parameter is to represent a stack element.

However, the existing algorithms cannot generate a generic fluent API. In this paper, a generic fluent API refers to a fluent API that provides generic methods, which are those including type parameters in their definitions. The Stream API in Java is an example of such a generic API; it provides generic methods such as `map(Function<? super T, ? extends R> mapper)`. A generic method uses a type parameter to check the *semantic* constraints of the API; for example, to check whether a correct value type is passed through a stream. However, in previous studies, the role of a type parameter was limited to the internal representation of a stack element of the parser built on the type system. The lack of support for generic methods is problematic when using generation algorithms for fluent APIs in real-world settings. Thus, adopting a type parameter is a crucial feature for using the API in a statically type-safe manner.

In this paper, we propose Protocool, a tool for generating a generic fluent API. Unlike the case in previous studies, the users of Protocool can include generic methods in the grammar definition. Our translation algorithm implemented for Protocool is modeled as the construction of deterministic finite automaton (DFA) with type parameter information. Each state of the DFA holds information regarding which type parameters are already bound in that state. The information is used to identify whether a method invocation in a chain newly binds a type to a type parameter or refers to a previously bound type. This identification is necessary because a type parameter in a chain is bound at a particular method invocation, and that bound type is referenced in the following method invocations. Our algorithm constructs the DFA by analyzing the binding times of type parameters and their propagation among the states in a DFA that is naively constructed according to the given grammar.

Although our algorithm receives regular grammar and constructs a DFA, the checking of context-free syntax can also be achieved by specifying the use of nested generics to represent a stack structure.

The remainder of this paper is organized as follows: Section 2 illustrates the problems of existing algorithms when generating a generic fluent API. Section 3 describes our translation algorithm from the given grammar to a safe fluent API. In section 4, we present use cases to demonstrate the ability and limitations of Protocool. Section 5 discusses related work, and section 6 concludes our proposal.

## 2 Motivation

As a motivating example, consider the creation of a generic fluent API for constructing an instance of Map<K, V> in Java. Our example API allows its users to construct a map with any key/value type by chaining the method invocations, as follows:

```
Map<Integer, String> map = OurAPI.newMap()
    .put(1, "foo") // put an entry with key = 1, value = foo
    .put(2, "bar") // put an entry with key = 2, value = bar
    .build();
```

In our API, the key and value types are inferred from the types of the argument provided to the first invocation of put(...) in the chain. Users cannot create an entry with an inconsistent key/value type. A type error is reported if such an entry is created:

```
OurAPI.newMap()
    .put(1, "foo") // key: Integer, value: String
    .put("bar", 2) // key: String, value: Integer <- Causes a type error!
    .build();
```

Our API also includes *syntactical* rules regarding the order of the method invocations in a chain. Users first need to invoke newMap() to begin a map construction. Thereafter, they can create entries by chaining an arbitrary number of put(...). They need to invoke build() to complete the construction and obtain an instance of Map<K, V>. The syntax described above can be summarized as follows, in the form of a regular expression:

$$\text{newMap()} \ \big(\text{put(K key, V value)}\big)* \ \text{build()}$$

The asterisk denotes zero or more occurrences of the preceding element, while K and V are type parameters that represent the key and value types, respectively.

Consider reporting the violation to the syntax as a type error so that users can identify their misuse at compile-time. This *safe* property can be achieved by setting the return type of each API method based on what the users can chain next. For example, a duplicate invocation of newMap() can be prevented by setting the return type of newMap() to a class providing only put(...):

```
OurAPI.newMap() // Returns a class providing only put(...)
    .newMap(); // This line causes a type error; Cannot resolve method 'newMap'
```

It is known that such a *safe* property can be achieved by: (1) building a state machine that accepts only a syntactically correct method sequence, and (2) encoding the machine into Java class definitions [3, 16, 29]. Recall that the syntax of our API is expressed by a regular expression. Therefore, a DFA is capable of recognizing the syntax. Figure 1 illustrates the DFA that accepts method sequences conforming to the syntax.

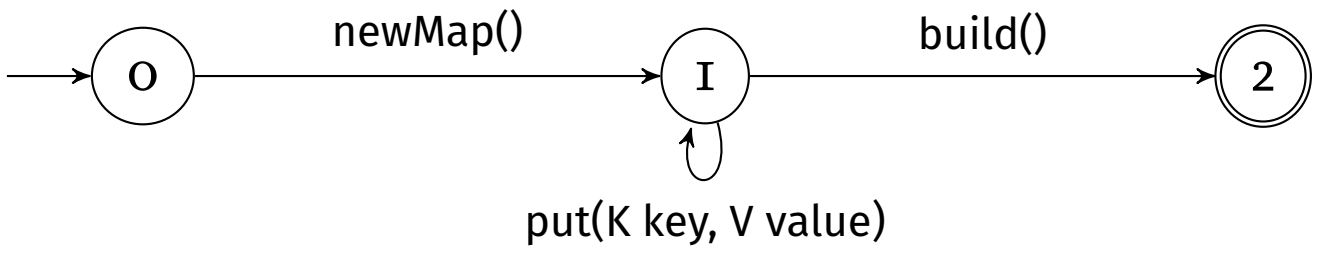

**Figure 1** DFA that accepts correct method sequences of OurAPI. The left-most state in the figure is the initial state of the DFA. The double circle represents an accepting state. The number in each state is simply an index of the state for later references.

A problem arises when encoding the DFA into class definitions. According to previous studies [3, 16, 29], we obtain the class definitions by encoding each state into a class, and encoding each transition into a method. Consider encoding the loop transition consuming put(...) into a method. A naive idea is to encode it into the following:

```
class State1<K, V> { // Corresponds to the state numbered with 1 in the DFA diagram.
    State1 put(K key, V value) { /* method body */ }
    ...
}
```

Unfortunately, this encoding does not generate our example API as expected. In this encoding, all invocations of put(...) refer to types that are bound to the type parameters K and V. However, no invocations of put(...) bind the argument types to K or V. Thus, the API generated by this encoding is broken. Another idea is to encode the loop transition, as follows:

```
class State1 {
    <K, V> State1 put(K key, V value) { /* method body */ }
    ...
}
```

Using this encoding, an invocation of put(...) binds the argument types to K and V. However, this encoding also does not generate our example API. In this encoding, K and V are bound at every invocation of put(...). As no invocations refer to a previously bound type, the map entry types become inconsistent.

The problem is that, in the state machine naively constructed from syntactical rules, it is not clear which type parameters are already bound in each state. A type parameter in a method chain is bound at the first method invocation that uses the type parameter. The successive method invocations refer to that bound type. In our API, the type parameters K and V are bound to the argument types provided to the first invocation of put(...). Successive invocations of put(...) refer to those bound types for their arguments. The binding rule of a type parameter in a chain is not specific to our API. For example, the type parameter in the Stream API is bound to a type, as described above:

```
Stream.of("a", "aa", "aaa") // Bind String to T of Stream<T>
    .filter(s -> s.length() > 1) // Refer to the bound type; The type of s is String
    .forEach(s -> System.out.println(s)); // Refer to the bound type; The type of s is String
```

**Listing 1** Specification of OurAPI

```
1 class OurAPI {
2     static Map<K, V> newMap() put(K key, V value)* build(); // Defines syntax
3     K;   V; // Define type parameters for this class
4 }
```

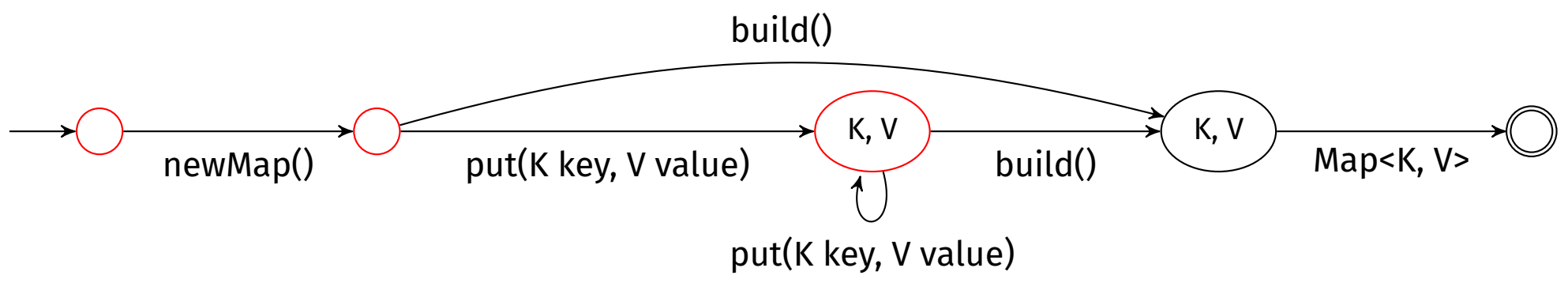

**Figure 2** DFA constructed from class declaration in listing 1

To generate a generic fluent API correctly, an algorithm needs to construct a state machine that knows which type parameters are bound in each state. If the bound type parameters in each state are clear, the encoding algorithm can identify whether or not a generated method newly binds types to type parameters.

## 3 Protocool

To address the problem mentioned in the previous section, we propose Protocool, a tool that can generate a generic fluent API.

Protocool receives the API specification written in Java-like syntax. Listing 1 illustrates the specification of OurAPI, which is the example API described in the previous section. The statement on line 2 in listing 1 is a chain declaration, which defines the syntactically correct chaining of the API methods. The keyword static merely indicates that the first method of a chain is a static method. The generated API is used as follows if we remove static:

```
// "new OurAPI().newMap()" instead of "OurAPI.newMap()"
Map<Integer, String> map = new OurAPI().newMap().put(1, "foo").build();
```

The statements on line 3 declare the type parameters used in the API.

According to the given specification, Protocool constructs a DFA in which each state is annotated with the type parameters bound in that state. Figure 2 illustrates the DFA constructed by Protocool from the specification in listing 1. The symbols indicated inside a state circle are the type parameters bound in that state. The DFA consumes a method or type at each step. It reaches an accepting state by consuming the method sequence defined in the chain declaration, and then by consuming the return type. A transition consuming a type identifies which type is instantiated by chaining methods. The construction of such a DFA consists of two steps. The first step is the naive construction of a DFA from the given syntactical rules. The second step is the modification of the naively constructed DFA. These two steps for constructing a DFA are described in section 3.1.

**Listing 2** Class definitions generated from DFA in figure 2

```
1  class OurAPI { // Corresponds to the initial state
2      static State1 newMap() { . . . }
3  }
4  class State1 { // Corresponds to the second state from the left
5      <K, V> State2<K, V> put(K key, V value) { . . . }
6      <K, V> Map<K, V> build() { . . . }
7  }
8  class State2<K, V> { // Corresponds to the third state from the left
9      State2<K, V> put(K key, V value) { . . . }
10     Map<K, V> build() { . . . }
11 }
```

Protocool generates a safe fluent API by encoding the constructed DFA into Java class definitions. Listing 2 illustrates the class definitions generated from the DFA in figure 2. A state is encoded into a class if the state does not have a transition consuming a type and the state is not an accepting state. In figure 2, only those states colored with red are encoded into classes, as the others have a transition consuming a type or are an accepting state. The initial state is encoded into a class with the same name as the class declaration. In our example, the initial state is encoded into a class named `OurAPI`, as indicated on line 1 in listing 2. Other states are encoded into classes named, for example, `StateN`. A transition is encoded into a method if the transition consumes a method. The return type of a method depends on the destination state of the original transition. If the destination state includes a transition consuming a type, the return type is that consumed type. In our example, the return type of `build()` is `Map<K, V>`, as indicated on line 6 and line 10. Otherwise, the return type is the class corresponding to the destination state. The generated method includes its type parameter declaration when the type parameters bound in the source state differ from those that are bound in the destination state. Note that, if no `put(...)` is invoked in a chain, `K` and `V` are inferred from the type information outside of the chain. For example, if the return value of `OurAP.newMap().build()` is assigned to a variable, `K` and `V` are inferred from the type of that variable:

```
// K and V are bound to Integer and String, respectively.
Map<Integer, String> m = OurAP.newMap().build();
```

The method bodies are omitted here. We describe their generation in section 3.2.

### 3.1 DFA construction

Suppose that a given specification has the following class declaration:

class $c$ {
    $r_1\ s_{1,1}\ s_{1,2}\ \ldots\ s_{1,l_1}$;
    $\ldots$;
    $r_n\ s_{n,2}\ s_{n,2}\ \ldots\ s_{n,l_n}$;
}

Here, $c$ is a class name, $r_i$ is a return type, and $s_{i,j}$ is a method.

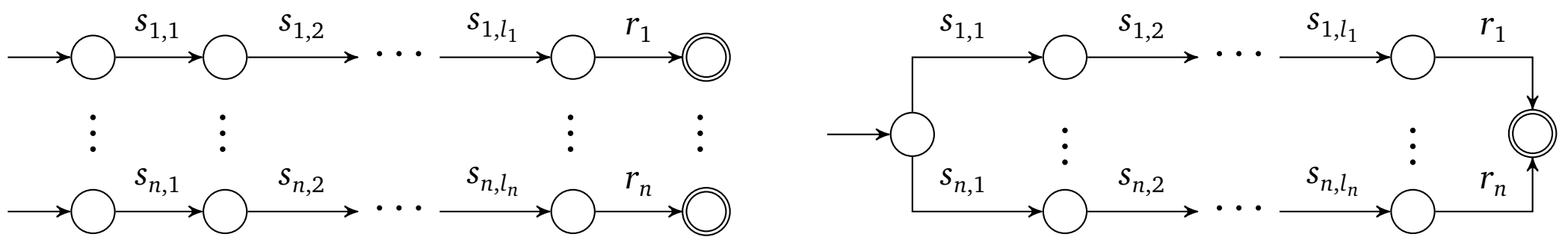

**(a)** Set of chain automata **(b)** NFA

**Figure 3** NFA construction

Protocool first constructs an NFA from the class declaration. It achieves this by combining a set of chain automata, each of which is constructed from a chain declaration. A chain automaton reaches an accepting state by consuming the method sequence of its corresponding chain declaration, and then by consuming the return type. Figure 3a illustrates the set of chain automata obtained from the above class declaration. Our algorithm constructs the NFA presented in figure 3b by merging the initial states and accepting states of the chain automata in figure 3a. Such an NFA is always constructed successfully from a given class declaration.

Protocool then converts the constructed NFA into a DFA using Brzozowski's algorithm [4]. To judge the equality between two transitions, our algorithm uses the function that takes two transitions as its input and returns a boolean value, as follows:

- returns true if both transitions consume methods and have the same signature;
- returns true if both transitions consume the same type; and
- returns false otherwise.

This conversion is necessary to generate a valid Java class definition, as non-determinism produces duplicate method definitions in a class. The conversion always succeeds because any NFA can be converted into a DFA.

Protocool finally analyzes the binding times of the type parameters to determine which type parameters are already bound in a state. It incrementally assigns a set of type parameters to each state of the DFA, and modifies the DFA constructed at the previous step during the analysis if necessary.

The algorithm for the binding time analysis can be described as follows. The algorithm first assigns the set of type parameters appearing in the class name to the initial state. In the example case, it assigns an empty set $\emptyset$ to the initial state, as illustrated in figure 4a, because `OurAPI` does not have any type parameters in its name. Suppose that a set of parameters $P_i$ is already assigned to a state $q_i$, and $q_i$ includes a transition $t : q_i \xrightarrow{s} q_j$. If any set has not been assigned to $q_j$, our algorithm assigns $P_i \cup \pi(s)$ to $q_j$, where $\pi(s)$ is a function to retrieve the set of type parameters appearing in $s$. In the example case, the algorithm assigns $\emptyset$ to the second state, as illustrated in figure 4b, because no set is yet assigned to the second state. If a set of parameters $P_j$ has already been assigned to $q_j$ and $P_j \neq P_i \cup \pi(s)$, our algorithm changes the destination of $t$ to a state $q'_j$ and assigns $P_i \cup \pi(s)$ to $q'_j$. In this case, $q'_j$ is a newly added state that is obtained by cloning $q_j$. The cloned state $q'_j$ has the same set of transitions as that of $q_j$. Figure 4c and figure 4d illustrate the cloning process in our example case. The algorithm continues to assign the bound type parameters to a state until all states are annotated with their bound type parameters.

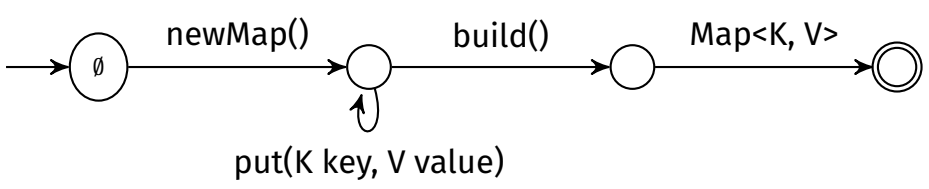

**(a)** Assign empty set to initial state

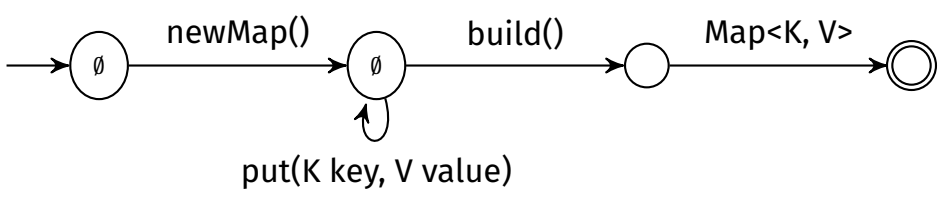

**(b)** Assign empty set to second state

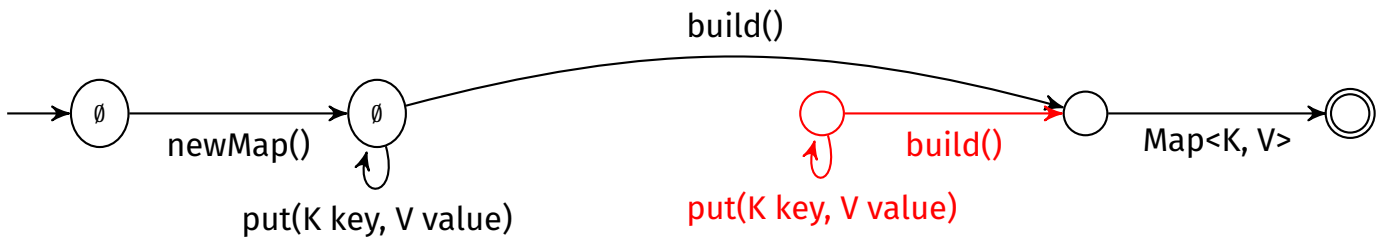

**(c)** Clone second state

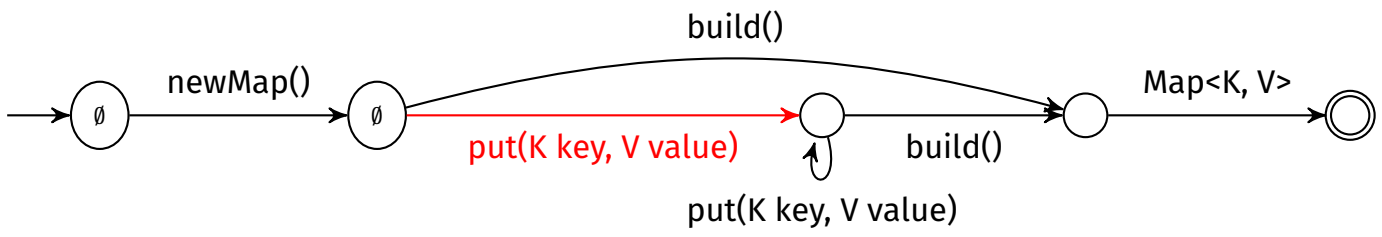

**(d)** Change loop transition destination

**Figure 4** Incremental assignment of type parameters

Algorithm 1 presents the pseudo-code of our algorithm. The function $\gamma(q, D)$ on line 18 is the function that clones a state $q$ in a DFA $D$. Because the cloning process does not add a transition, but only changes its destination, the modified automaton is still finite and deterministic. The binding time analysis does not repeat infinitely. Only a finite number of type parameters exist in a class declaration. A state is mapped to a subset of the set of those type parameters. Therefore, the number of states is also finite.

### 3.2 Bodies of generated methods

Protocool generates method bodies, as do the existing fluent API generators [17, 30]. The generation of method bodies helps generator users to implement the actions of a generated API. Without the body generation, users need to deal with the laborious task of restoring previously implemented actions when regenerating the API to change the API specification.

The generated method bodies construct a tree that represents a method chain composed by the API user. For example, the tree illustrated in figure 5 can be constructed from the following chain:

```
OurAPI.newMap().put(1, "foo").put(2, "bar").build();
```

Each node of the tree represents either an object construction or a method invocation. In figure 5, the root node `Object_Map` represents an object construction, while the child nodes such as `Method_newMap` represent a method invocation. The child nodes of an object construction node are method invocation nodes, each of which represents a method that is invoked to construct the object. The child nodes of a method invocation node are the arguments passed to that method.

Library developers (that is, Protocool users) can access the tree by specifying a tree evaluator through a return clause, as follows:

**Algorithm 1** Binding time analysis of type parameters

**Input:** A DFA $D$
**Output:** A DFA with type parameter information

$Q \leftarrow$ An empty queue
$q_0 \leftarrow$ The initial state of $D$
Assign the set of type parameters of the declared class to $q_0$
Enqueue all transitions outgoing from $q_0$ to $Q$
**while** $Q$ is not empty **do**
  $(t : q_i \xrightarrow{s} q_j) \leftarrow$ Dequeue from $Q$
  **if** $s$ is a method **then**
    $P_i \leftarrow$ The set assigned to $q_i$
    **if** No set is assigned to $q_j$ **then**
      Assign $P_i \cup \pi(s)$ to $q_j$
      Enqueue all transitions outgoing from $q_j$ to $Q$
    **else**
      $P_j \leftarrow$ The set assigned to $q_j$
      **if** $P_j \neq P_i \cup \pi(s)$ **then**
        **if** $\pi(s) = \emptyset$ **then**
          Assign $P_i \cup \pi(s)$ to $q_j$
        **else**
          $q'_j \leftarrow \gamma(q_j, D)$
          Assign $P_i \cup \pi(s)$ to $q'_j$
          Change the destination of $t$ to $q'_j$
          Enqueue all transitions outgoing from $q'_j$ to $Q$
        **end if**
      **end if**
    **end if**
  **end if**
**end while**

```
class OurAPI {
    static Map<K, V> newMap() put(K key, V value)* build() return Evaluator.buildMap;
    K; V;
}
```

In this case, Evaluator.buildMap is a static method defined by hand outside of the generated code. The constructed tree is passed to the static method placed after the keyword return:

```
Map<K, V> build() {
    // Create and store a new method node
    Method_build method_build = new Method_build();
    ...
    // Create a new object construction node
    Object_Map<K, V> object_map = new Object_Map<K, V>();
    ...
    // Pass the tree to the evaluator method and return the return value of the evaluator
    return Evaluator.buildMap(object_map);
}
```

Using this design, library developers can implement the semantics of the generated API separately from the generated code. The generated tree nodes support the visitor pattern. Listing 3 and listing 4 present example implementations of the semantics that construct a HashMap<K, V> instance using the visitor pattern.

## 3.3 Specification validation

Protocool throws an error and does not generate Java class definitions when it determines that a state with a type-consuming transition also has another transition. This is because the specification producing such a DFA cannot be translated into valid Java class definitions, or it is translated into unexpected Java classes.

### 3.3.1 State with multiple type-consuming transitions

Figure 6a presents an example specification that produces a state with multiple type-consuming transitions. It states that the users of the generated API can write both of the following:

```
List<String> list = Collection.of("foo").create();
Set<String> set = Collection.of("bar").create();
```

Figure 6b illustrates the DFA constructed from the specification presented in figure 6a. The state colored in red has two transitions that consume types.

Although Protocool can construct a DFA from the specification in figure 6a, this DFA cannot be translated into valid Java classes. The second state is encoded into the following Java class:

```
class State1<E> {
  List<E> build() { ... }
  Set<E> build() { ... }
}
```

However, in Java, methods with the same signature need to return the same type. Protocool throws an error to prevent the generation of a broken API.

### 3.3.2 State with both type-consuming and method-consuming transitions

Figure 7a presents an example specification that produces a state with both type-consuming and method-consuming transitions. It states that the users can write the following:

```
Map<String, String> map = StrMapBuilder.newMap().add("foo", "bar").add("bar", "baz");
```

Figure 7b illustrates the DFA constructed from the specification presented in figure 7a. The state colored in red has both type-consuming and method-consuming transitions.

Although a DFA is constructed by Protocool, the encoding of this DFA is problematic. The initial state is encoded into the following class, as the destination of the build() transition has a type-consuming transition:

```
class StrMapBuilder {
    static Map<String, String> newMap();
}
```

However, the users of the generated API cannot chain the method add(...), because Map<String, String> in Java does not provide add(...):

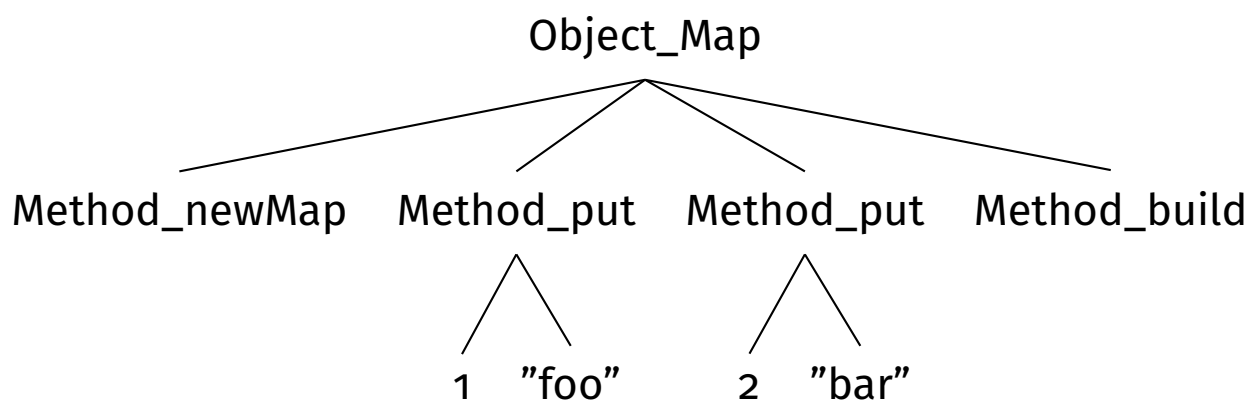

**Figure 5** Tree construction in generated library

**Listing 3** Handwritten evaluator implementation

```
class Evaluator {
  static <K, V> Map<K, V> buildMap(Object_Map<K, V> node) {
    BuildMapVisitor<K, V> visitor = new BuildMapVisitor<K, V>();
    visitor.visit(node);
    return visitor.map;
  }
}
```

**Listing 4** Handwritten visitor implementation

```
class BuildMapVisitor<K, V> extends Visitor {
  Map<K, V> map = new HashMap();
  void visitMethod_put(Method_put<K, V> node) {
    map.put(node.key, node.value);
  }
  void visitObject_Map(Object_Map<K, V> node) {
    super.visitConstruction_Layer(node); // visit child nodes
  }
  void visitMethod_newMap(Method_newMap node) {} // Do nothing
  void visitMethod_build(Method_build node) {} // Do nothing
}
```

```
class SingletonCollection {
    static List<E> of(E elem) build();
    static Set<E> of(E elem) build();
    E;
}
```

**(a)** Specification

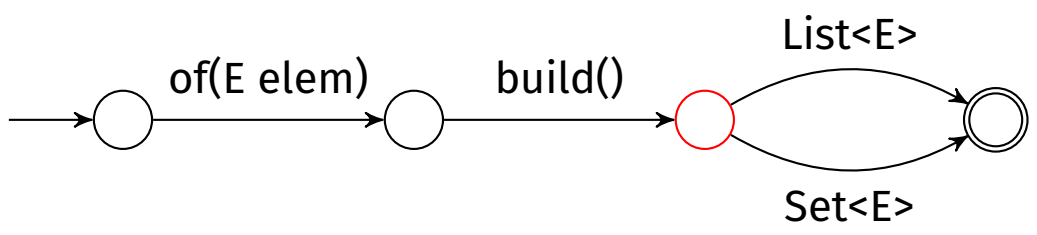

**(b)** DFA

**Figure 6** Invalid specification producing state with multiple type-consuming transitions

```
class StrMapBuilder {
    static Map<String, String>
      newMap() add(String k, String v)*;
}
```

**(a)** Specification

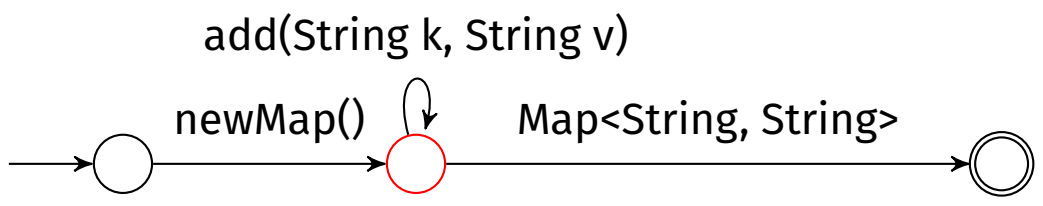

**(b)** DFA

**Figure 7** Invalid specification producing state with both type-consuming and method-consuming transitions

```
Map<String, String> map = StrMapBuilder
    .newMap() // Returns Map<String, String>
    .add("foo", "bar"); // Type error! Cannot resolve method 'add'
```

Protocool throws an error and do not generate code to prevent the generated API from being an unexpected API.

## 4 Use case

In this section, we illustrate the ability and limitations, as well as several features that are introduced for the practical applicability of Protocool through the generation of three APIs. The concrete syntax and semantics of our API specification language are presented in listing 5.

**Listing 5** Syntax of our API specification language. Parentheses are used to group elements. An asterisk represents zero or more occurrences, a plus sign represents one or more occurrences, and a question sign represents zero or one occurrence of the preceding element. A colon is used to define a lexical token. The left-hand side of a colon is the name of the token, while the right-hand side is the regular expression that the token should follow.

```
<spec> → <class>+ ;
<class> → <class-head> <class-body> ;
<class-head> → "class" NAME <type-param-list>? ;
<type-param-list> → "<" <type-param> ( "," <type-param> )* ">" ;
<class-body> → "{" <chain-or-type-param>* "}" ;
<chain-or-type-param> → ( <chain> | <type-param> ) ";" ;
<type-param> → NAME <type-param-bound>? ;
<type-param-bound> → "extends" <type-ref-list> ; # Defines upper bound of a parameter
<type-ref-list> → <type-ref> ( "," <type-ref> )* ;
<chain> → "static"? <type-ref> <chain-expr> <tree-eval>? ;
<chain-expr> → <chain-term> ( "|" <chain-term> )* ;
<chain-term> → <chain-fact>+ ;
<chain-fact> → <chain-elem> ( "?" | "*" | "+" )? ;
<chain-elem> → <method> | "(" <chain-expr> ")" ;
<method> → NAME "(" <method-param-list>? ")" <method-action>? ;
<method-param-list> → <method-param> ( "," <method-param> )* ;
<method-param> → <type-ref> "..."? NAME ;
<method-action> → "{" <qual-name> ";" "}" ;
<type-ref> → <qual-name> ( "<" <type-ref-list> ">" )? "[]"* ;
<tree-eval> → "return" <qual-name> ; # Specifies a tree evaluator method
<qual-name> → NAME ( "." NAME )* ;
NAME : [a-zA-Z_][a-zA-Z0-9_]* ;
```

**Listing 6** Specification of our matrix library

```
1 class MatrixBuilder {
2   ROW extends Size; // Size is upper bound of parameter ROW
3   COL extends Size;
4   static IntMat<ROW, COL> randInt() row(ROW row) col(COL col);
5   static FltMat<ROW, COL> randFlt() row(ROW row) col(COL col);
6 }
7 class IntMat<ROW extends Size, COL extends Size> {
8   NEW_COL extends Size;
9   IntMat<ROW, COL> plus(IntMat<R, COL> m);
10  FltMat<ROW, COL> plus(FltMat<R, COL> m);
11  IntMat<ROW, NEW_COL> mult(IntMat<COL, NEW_COL> m);
12  FltMat<ROW, NEW_COL> mult(FltMat<COL, NEW_COL> m);
13  int[][] toArray() return Evaluator.toIntArray;
14 }
15 class FltMat<ROW extends Size, COL extends Size> {
16  NEW_COL extends Size;
17  FltMat<ROW, COL> plus(IntMat<ROW, COL> m);
18  FltMat<ROW, COL> plus(FltMat<ROW, COL> m);
19  FltMat<ROW, NEW_COL> mult(IntMat<COL, NEW_COL> m);
20  FltMat<ROW, NEW_COL> mult(FltMat<COL, NEW_COL> m);
21  float[][] toArray() return Evaluator.toFloatArray;
22 }
```

### 4.1 Matrix computation API – checking complex protocol using type parameters

Support for type parameters enables examination of a relatively complex API protocol. As an example, we demonstrate the generation of a matrix computation API that reports an incompatible computation as a type error.

Our matrix computation API provides two classes, namely IntMat and FltMat, which represent an integer matrix and a float matrix, respectively. The API supports matrix addition and multiplication. Only a matrix computation between integer matrices returns an integer matrix. Other computation, such as addition between an integer matrix and a float matrix, returns a float matrix.

Listing 6 illustrates the specification of our matrix computation API. The type parameters ROW and COL are the row and column sizes of a matrix, respectively. The boundings for these parameters are written in a similar manner to that of Java on line 7 and line 8 in listing 6. (The keyword extends defines the upper bound of a type parameter.) The boundary type Size is a type that is defined manually, as follows:

```
abstract class Size { abstract int getIntVal(); }
```

The API users can define any matrix size by subclassing Size outside of the Protocool-generated code. For example, if they use 128-by-256 matrices and 256-by-128 matrices in their computation, they need to define the two classes Size128 and Size256. The abstract method getIntVal is used in the tree evaluator to obtain the integer value represented by a concrete Size class.

The users of our matrix computation API proceed as follows:

```
Size128 size128 = new Size128(); // class Size128 extends Size { ... }
Size256 size256 = new Size256(); // class Size256 extends Size { ... }

FltMat<Size128, Size128> matrix1 = MatrixBuilder.randFlt().row(size128).col(size128);
IntMat<Size128, Size256> matrix2 = MatrixBuilder.randInt().row(size128).col(size256);
FltMat<Size128, Size256> matrix3 = matrix1.mult(matrix2);
```

The following statements throw type errors, as they do not conform to the protocol of our API:

```
FltMat<Size128, Size128> matrix1 = MatrixBuilder.randFlt().row(size128).col(size128);
IntMat<Size128, Size256> matrix2 = MatrixBuilder.randInt().row(size128).col(size256);

// Cause type errors (incompatible sizes)
matrix2.mult(matrix1); // [128, 128] * [256, 128]
matrix1.plus(matrix2); // [128, 256] + [128, 128]

// Cause type errors (incompatible element types)
IntMat f2x3 = matrix1.mult(matrix2); // FltMat * IntMat returns FltMat
```

When using the API in Scala 2.13, users can avoid defining a custom-sized class by using literal singleton types [21], which allows programmers to use a literal value as a type. To achieve this, the following helper function and class need to be defined by hand:

```
def size[T <: Singleton](v: T): SizeForScala[T] {
  new SizeForScala[T](t)
}
class SizeForScala[T <: Singleton](v: T) extends Size {
    override def getIntVal: Int = v.asInstanceOf[Int]
}
```

The function size creates an instance of SizeForScala[T], which extends the abstract class Size illustrated above. As the parameter T is bounded to Singleton, the type of v is inferred as a literal singleton type. For example, the return type of size(100) is the literal type 100, which is a subclass of scala.Int. Using the helper function and class illustrated above, users can write their computation as follows:

```
val matrix1 = MatrixBuilder.randFlt().row(size(128)).col(size(128));
val matrix2 = MatrixBuilder.randInt().row(size(128)).col(size(256));
val matrix3 = matrix1.mult(matrix2);
matrix2.mult(matrix1); // Causes a type error
```

### 4.2 Itemized document API – checking context-free grammar

As Protocool allows its users to specify the manner in which to use type parameters in their API, they can use type parameters to check context-free grammar. As an example of APIs with context-free grammar, consider an API that emulates itemization of LaTeX, as follows:

```
begin() // \begin{itemize}
    .item("Item A") // \item Item A
```

```
    .begin() // \begin{itemize}
        .item("Item A.1") // \item Item A.1
    .end() // \end{itemize}
.end() // \end{itemize}
.asTeXStr();
```

The API requires its users to invoke only a pair of begin() and end() in a chain. A type error occurs when the API users attempt to obtain a string from the itemization with an unbalanced invocation of begin() and end():

```
begin()
    .item("Item A")
    .begin()
        .item("Item A.1")
    .end()
.asTeXStr(); // Causes a type error; Cannot resolve method 'toTeXStr'
```

In the API, the item type is inferred from the first argument provided to item(...). A type error occurs when users input an item with an inconsistent type:

```
begin()
    .item(100)
    .item("200") // Causes a type error
.end().asTeXStr();
```

The itemized document API described above can be generated from the specification illustrated in listing 7. The API uses the first type parameter to represent the stack of a pushdown automaton, as proposed in the literature [23]. The second type parameter represents the type of the items in the document.

Although it is possible to check context-free rules in the Protocool-generated API, it is often excessively tedious to specify such checking in the specification. To achieve such checking, Protocool users need to encode a pushdown automaton into the specification manually. The encoding of a pushdown automaton has been automated in previous studies.

However, this limitation will not degrade the practical applicability of Protocool, because context-free rules are not common in API design. The nesting structure is

**Listing 7** Specification for itemized document API

```
class API {
    ITEM;
    static Nested<EndOfDoc, ITEM> begin(ITEM item) ;
}
class Nested<X, ITEM> {
    Nested<Nested<X, ITEM>, ITEM> begin(ITEM item) ;
    X end(ITEM item) return Evaluator.end ;
}
class EndOfDoc {
    String asTeXStr();
}
```

**Listing 8** Specification of subset of AssertJ

```
class Assertions {
    PredicateAssert assertThat(String s);
}
class PredicateAssert {
    PredicateAssert startsWith(String s) { Action.startsWith; }
    PredicateAssert endsWith(String s) { Action.endsWith; }
}
```

often emulated with a feature that is available in the host language, such as method invocation syntax:

```
// Nesting is emulated by passing subchains to their parent chain
// With varargs:
itemize(item("Item A"), itemize(item("Item B"))).asTeXStr();
// Without varargs:
itemize().elem(item("Item A")).elem(itemize(item("Item B"))).asTeXStr();
```

This type of subchaining technique is frequently used in a real-world library such as j2html.[1] The problem with techniques using variable-length arguments is that all arguments need to be of the same type. This problem does not occur when using techniques without variable-length arguments. However, this requires somewhat redundant invocations of elem(...). Finding a succinct emulation with the ability to take different argument types can be investigated in future work. In relatively new languages, special syntactic sugars are provided that can be used to emulate nesting structures [7, 8].

### 4.3 Assertion API – shallow embedding

As described in section 3.2, the semantics of a Protocool-generated API are designed to be added by creating a tree evaluator. This style is known as the deep embedding style [15]. In this style, the execution is postponed until the API user invokes a tree evaluation method, such as build() in OurAPI. However, an API often does not provide such a tree evaluation method. AssertJ [2] is an example of such an API. It does not require its users to invoke a method such as runAssertions() at the end of the chain: every method call immediately runs an assertion.

```
import static org.assertj.core.api.Assertions.assertThat;
String str = . . . ;
assertThat(str).startsWith("A") // Runs assertion immediately here
               .endsWith("Z"); // Runs assertion immediately here
```

This style of implementing semantics is known as the shallow embedding style [15].

Protocool supports the shallow embedding style. Listing 8 illustrates the specification of the subset of AssertJ that uses the feature for the shallow embedding style. The {} block written after a method specifies an action invoked in that method. Protocool inserts the action directly before the return statement in the generated method body, as follows:

```
PredicateAssert startsWith(String s) {
  Object_PredicateAssert node = . . . ; // Tree construction
  Action.startsWith(node); // Inserted action
  return new PredicateAssert(node);
}
```

[1] https://j2html.com (Visited on 2020-01-26).
[2] http://joel-costigliola.github.io/assertj (Visited on 2020-01-26).

The constructed tree is provided to the inserted action. Unlike the return clause described in section 3.2, the action return value is discarded.

## 5 Related work

### 5.1 Typestate analysis

Typestate analysis [27] is a form of program analysis that verifies whether an operation sequence performed on an object follows specified rules (or protocols). Objects with typestates occur quite frequently. According to the literature [2], 7.2% of Java types define protocols.

Various techniques have been proposed to realize typestate analysis. Plaid [28] is a language that inherently provides features for typestate analysis. Because typestate analysis is a type of static code analysis, it can be achieved by using general-purpose code analyzers such as FindBugs,[3] PMD,[4] and QL [1]. Techniques for mining typestate specifications have also been studied to overcome the difficulty of completely defining the specification by hand [10, 13, 14, 25].

Generating a fluent API can be regarded as a technique for realizing typestate analysis with the type system of a language. It encodes each state of a type into a concrete type definition of a target language. The validity of the operation sequences is checked by the type system of that language. Such a generative approach causes two problems that do not occur when using external analyzers. Firstly, strange names are given to intermediate states, which may confuse API users. Secondly, performance deterioration may occur owing to the increase in the type definitions. These problems have not been studied yet as far as we know and their investigation will form part of our future work. Although the generative approach suffers from these disadvantages, it offers advantages that are not immediately provided by external analyzers. It aids API users in that a method completion system becomes state aware [23, 29], and it is also beneficial to library developers. As the type checker rejects code violating the protocol, developers do not need to add the implementation for handling such cases.

[3] http://findbugs.sourceforge.net. (Visited on 2020-01-26.)
[4] https://pmd.github.io. (Visited on 2020-01-26.)

### 5.2 Fluent API generators

Numerous tools and algorithms have been proposed to translate BNF-style grammar into a safe fluent API.

TS4J [3] is a tool for generating a safe fluent API with regular grammar. It constructs a DFA from a given grammar, but does not analyze the binding times of the type parameters and their propagation among the states. The tool demonstrated in the literature [5] also constructs a DFA to generate a fluent API, but also does not analyze the binding time.

EriLex [29] is a tool for translating LL grammar into a safe fluent API in Scala. It constructs a real-time deterministic pushdown automaton (RDPDA), and then encodes that automaton into Scala classes. Although the output is a Scala API, the technique used in EriLex can be exported into Java, because the generated API does not use Scala-specific features. Silverchain [23] is a tool for generating a Java fluent API that supports subchaining. As with EriLex, the translation algorithm of Silverchain constructs an RDPDA; therefore, the supported grammar class is LL grammar.

Gil and Levy proposed an algorithm to translate LR grammar into a safe fluent API in Java [16, 22]. Their algorithm constructs a jump-stack single-state real-time deterministic pushdown automaton [9], which is a machine that is equivalent to an LR parser. Thereafter, it encodes that automaton into Java classes. However, the compilation time of a chain composed using the generated API increases exponentially with the chain length in the worst case. Gil and Roth proposed another algorithm for LR grammar to overcome the exponential growth [17]. Their newer technique uses tree encoding to emulate a deterministic pushdown automaton on Java's type system.

TypelevelLR [30] is a tool for translating LR grammar into a safe fluent API in Scala, Haskell, and C++. The APIs in Scala and Haskell use type classes to encode $\epsilon$ transitions of an LR parser into type definitions. The API in C++ uses C++ templates to encode the $\epsilon$ transitions.

Grigore demonstrated that Java's type system is Turing complete [18]. Using this result, he illustrated that a CYK parser [6, 20, 31] can be constructed using Java types. Although he discovered the theoretical upper bounds of grammar classes that can be checked by Java's type system, his technique requires a large memory size. This is because it builds the parser on top of the Turing machine implemented by the Java types.

As demonstrated in section 4.2, Protocool users can generate an API with context-free syntax checking. The users can generate such an API by explicitly using type parameters to represent a stack structure. This capability is owing to the support of generic methods (binding time analysis of type parameters) in our algorithm. Furthermore, the support of generic methods enables checking of the semantic (non-syntactical) constraints of the generated API, as demonstrated in section 4.1. Such API constraints have not been thoroughly considered in previous generators.

### 5.3 Domain-specific language

Creating fluent APIs is a popular means of implementing domain-specific languages inside a general-purpose language. However, it is not the only option that introduces a domain-specific notation into general-purpose programs, for which syntax extension is a well-known solution. SugarJ [11] provides a method for extending Java syntax. ProteaJ [19], Wyvern [24], and Honu [26] are programming languages that natively support syntax extension. Using these syntax extension mechanisms, domain-specific notation can be embedded as is. However, when creating a fluent API, such notation needs to be transformed into a method chain that differs slightly from the original notation. Despite this drawback, creating a fluent API still offers an advantage: a fluent API only requires method invocation syntax such as obj.method(...), which is offered by most object-oriented languages.

## 6 Conclusions

We have proposed Protocool, a tool for generating a generic fluent API in Java. The contribution of this paper is the development of the translation algorithm implemented in Protocool. Unlike the methods of previous studies, our algorithm analyzes the binding times of type parameters in a method chain. The analysis is the key technique enabling the generation of a generic fluent API. Support for generic methods is essential for introducing safe fluent API generation into the real world, as generic methods are frequently used to make an API statically type safe. It is possible to check context-free rules in the Protocool-generated API, as the use of type parameters is completely the decision of the Protocool users.

Further investigation into experience with using Protocool will be our primary future work. In particular, studying the effects on library user experiences caused by strange type names, and the time and space overheads caused by the code bloat, will be useful for discussing the practical applicability of Protocool and other fluent API generators.

## About the authors

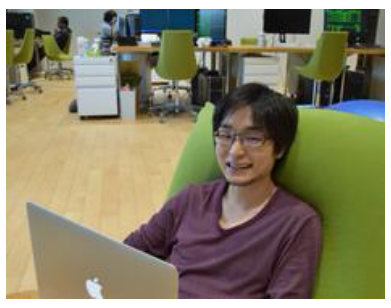

**Tomoki Nakamaru** is a Ph.D student at The University of Tokyo. His research interests include programming languages, program analysis, and repository mining. Contact him at nakamaru@csg.ci.i.u-tokyo.ac.jp

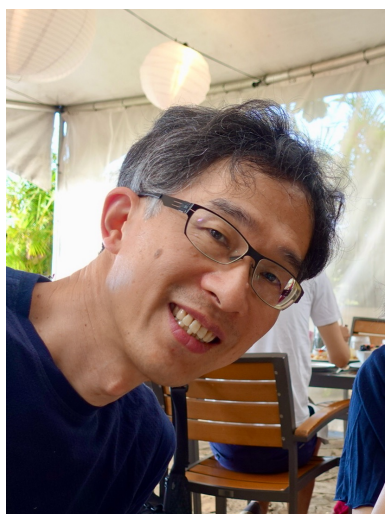

**Shigeru Chiba** is Professor at The University of Tokyo. Shigeru's research focuses on programming language design, implementation, tools, and libraries. He has been publishing papers on reflection and meta-programming, aspect-oriented programming, and embedded domain specific languages at OOPSLA, ECOOP, and other conferences. He has also served as a program committee member at those conferences. He received 2012 IBM Faculty Award. He is also a primary developer of Javassist, which is a Java bytecode engineering toolkit widely used in industry and academia. He received his PhD in computer science from the University of Tokyo. Contact him at chiba@acm.org